# Functionally Reconfigurable Integrated Structure of Fabry-Perot Antenna and Wideband Liquid Absorber

Yukun Zou, Xiangkun Kong, *Member, IEEE*, Zuwei Cao, Xinyu Zhang, He Wang, Xuemeng Wang, Yongjiu Zhao

*Abstract*—This paper proposes a functionally reconfigurable integrated structure of a Fabry-Perot (FP) antenna and wideband liquid absorber. The partial reflecting surface (PRS) is elaborately tailored to serve as both a component of the FP antenna and the metal ground of the broadband liquid absorber. Then the integrated structure is realized by combining the FP antenna with the liquid absorber. The PRS is composed of patches on the top layer of the substrate and the square loop on the bottom and the microstrip patch antenna is used as a source. The liquid absorber is composed of a 3-D printed container, 45% ethanol layer, and a metal ground. By this design, the structure can serve as the FP antenna when the ethanol is extracted or the liquid absorber when the ethanol is injected. This can be used to switch between detection and stealth mode which can better adapt to the complex electromagnetic environment. The simulation results show that the gain of the antenna reaches to 19.7 dBi and the operating band is 100 MHz when the ethanol is extracted. When the ethanol is injected, the absorption band of the absorber ranges from 4 GHz to 20 GHz and the absorption bandwidth is over 133%. The monostatic RCS reduction bands of the structure with ethanol range from 4 GHz to 20GHz and the average RCS reduction is 28.4 dB. The measured and simulated results agree well.

*Index Terms*—FP antenna, reconfigurable, liquid absorber, radar cross section (RCS).

## I. INTRODUCTION

FABRY-PEROT (FP) antenna which is also known as partially reflecting surface (PRS) antennas, 2-D leaky-wave antennas have attracted more and more attention in stealth platforms, sensing systems, and satellite communications because of their simple configurations and high directivity [1]-[4]. With the development of radar detection technologies, it is important to improve the stealth properties of antennas which play a key role in radar systems in the military background. The RCS is an evaluation criterion of stealth properties. Therefore, it is necessary to reduce the RCS of FP antennas.

Recently, many types of research about the RCS reduction of FP antennas have been published. Loading absorbing structures (AS) is one of the effective ways to reduce the RCS of FP antennas. Through the integrated design of AS and PRS [5]-[8], the incident waves will be absorbed so that the RCSs of antennas will be reduced. Despite the AS can effectively reduce the RCS of FP antennas, the gain of antennas is often affected. To maintain the gain of F-P antennas, chessboard-layout metasurfaces are employed to reduce the RCS of FP antennas [9]-[12]. By locating two kinds of artificial magnetic conductors (AMC) on the top of PRS, the RCS of the antenna will be reduced through the phase cancellation between the incident and reflected waves. Another way to reduce the RCS of the FP antenna is to apply the polarization rotation metasurfaces [13]-[15] to the FP antenna. In this way, the co-polarization waves will be transformed into cross-polarization waves so that the RCS of the antenna will be reduced. Coding metasurfaces [16] have also been applied to the RCS reduction of FP antennas. The incident waves will be scattered in all other directions by encoding techniques.

The technologies proposed above are all based on inactive material. The properties of the antenna keep unchanged once processed. Meanwhile, this can't adapt to the complicated electromagnetic environment. So far, many liquid materials have been applied to absorbers [17]-[18], antennas [19]-[20], frequency selective surfaces [21]-[22] and so on because of their wide absorption band, transparency, reconfigurable properties, low-RCS, and low price. To overcome the challenges of reconfigurable properties of FP antennas. This paper proposes a functionally reconfigurable integrated structure of FP antenna and wideband liquid absorber. The structure acts as an FP antenna when the ethanol is extracted while it acts as the absorber when the ethanol is injected. So that the structure can work between the radiation and stealth modes. The gain of the antenna reaches 19.7 dBi when the ethanol is extracted. When the ethanol is injected, the absorption band of the absorber ranges from 4 GHz to 20 GHz.

This paper will go as follows. Section II introduces the design and geometry of the FP antenna and the absorber. Section III presents the geometry and properties of the integrated structure. Finally, Section IV concludes the paper.

## II. DESIGN OF THE PRS AND THE ABSORBER

### A. Design of the FP antenna

The unit cell of the PRS is shown in Fig. 1(a), it is composed of the top metal patch, the F4B dielectric substrate ($\varepsilon_r$ = 2.2, tan$\delta$

This work was supported in part by the National Natural Science Foundation of China under Grant 62071227, in part by the Natural Science Foundation of Jiangsu Province of China under Grant BK20201289, in part by the Postgraduate Research & Practice Innovation Program of Jiangsu Province under Grant SJCX20_0070, in part by the Fundamental Research Funds for the Central Universities under Grant XCXJH20210402.

The authors are with the College of Electronic and Information Engineering, Nanjing University of Aeronautics and Astronautics, Nanjing, 211106, China (e-mail: xkkong@nuaa.edu.cn).

= 0.0009) and the bottom metal loop. The PRS has two functions: 1) acts as the metal ground of the absorber; 2) forms an FP cavity with the source antenna floor. The FP antenna consists of the PRS and the microstrip patch antenna which is used as a source, as shown in Fig. 2(b).

Although it can improve the absorption property of the absorber with a high reflection magnitude of the PRS, the gain of the antenna will decrease. Therefore, it must make a compromise between the absorption property of the absorber and the gain of the antenna. As shown in Fig. 2(a), with the increase of the width of the top patch $w_1$, the reflection magnitude of the PRS at 10 GHz will increase. However, the gain of the antenna at 10 GHz decreases as proposed forward. Last, the intersection of two curves of the reflection magnitude of PRS and the gain of the antenna has been chosen.

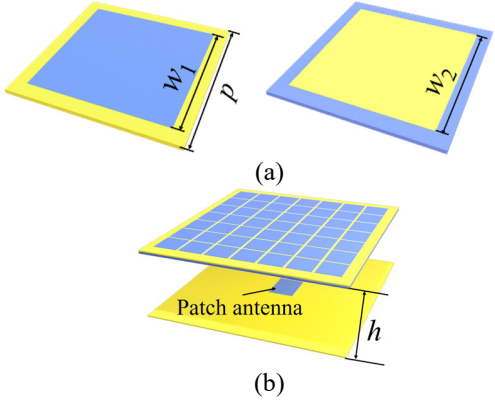

**Fig. 1.** (a) Unit cell of the PRS (b) 3D structure of the FP antenna.

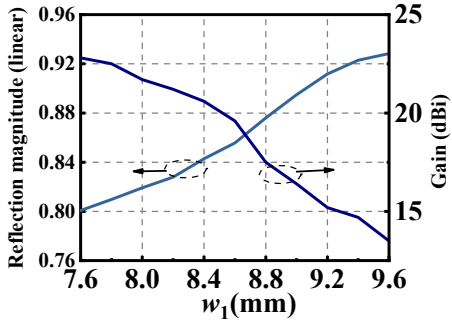

**Fig. 2.** Reflection magnitude of the PRS and the gain of the antenna with different $w_1$ at 10 GHz.

The operating frequency of the FP antenna is designed at 10 GHz, and the height of the FP resonant cavity is designed by the following equation [1], [24]:

$$h = \frac{\lambda}{4\pi}(\varphi_1 - \varphi_2 - 2N\pi), N = 0, 1, 2... \quad (1)$$

where $\varphi_1$ and $\varphi_2$ are the reflection phase of PRS and the metal ground, respectively. $\lambda$ is the wavelength of operating frequency. $N$ is the resonance mode number of the FP resonant cavity. By equation (1), the calculated value of the height of the FP resonant cavity $h$ = 14.2 mm. The optimized parameters are as follows: $w_1$= 8.6 mm, $p$ = 12 mm, $w_2$ = 8.6 mm, $h_1$ = 2 mm.

*B. Design of the liquid absorber*

As the finish of the PRS, the metal ground of the liquid absorber has been fixed. To expand the low-frequency absorption bandwidth of the liquid absorber under the fixed metal ground, two shapes of liquid absorbers are compared. One of them is a truncated cone-shaped structure and the other is a prismatic-shaped structure. The unit cell of the liquid absorber is both consisted of three layers: the 3d container layer ($\varepsilon_r$ = 2.8, tan$\delta$ = 0.0318), the water layer and the metal ground. The only difference is their shape. As shown in Fig. 3(a), keep the bottom diameter and the height of the truncated absorber unchanged, the first resonant frequency of the truncated cone-shaped absorber will go lower frequency with the increase of the top diameter $d_1$. Similarly, the first resonant frequency of the prismatic absorber will also go lower frequency with the increase of the top width $W_1$. However, compared to the truncated absorber, the first resonant frequency of the prismatic absorber is lower under a similar size.

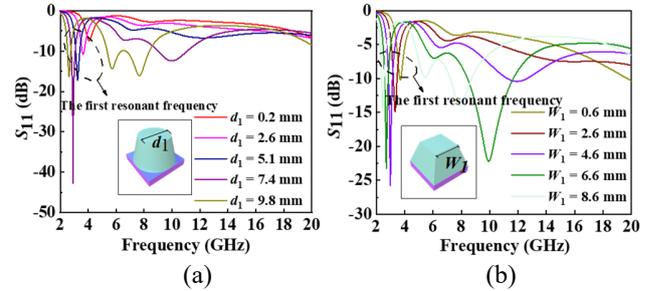

**Fig. 3.** (a) $S_{11}$ of the truncated cone-shaped liquid absorber with different top diameters $d_1$ (b) $S_{11}$ of the prismatic-shaped liquid absorber with different widths of top bottom $W_1$.

To verify this conclusion, the Mie resonance theory has been introduced. According to [23], the first resonant frequency can be calculated by the following formulas:

$$f = \frac{\varphi c}{2\pi r \sqrt{\varepsilon_p \mu_p}} \quad (2)$$

$$F(\varphi) = \frac{2(\sin\varphi - \varphi\cos\varphi)}{(\varphi^2 - 1)\sin\varphi + \varphi\cos\varphi} \quad (3)$$

$$r = k_0 \times (\frac{4}{3})^{k_1(h-n)} \quad (4)$$

where $f$ is the resonant frequency for the first-order Mie resonance, $r$ is the radius of the particle. c is the speed of light under vacuum, $\varepsilon_p$ and $\mu_p$ are the permittivity and permeability of material of particle, respectively. $F(\varphi)$ is defined as the resonant function related to $\varphi$ and $\varphi$ is approximately equal to $\pi$ under the first-order Mie resonance. In [21], a modified formula (3) has been given in the case of the water block. However, this doesn't apply to our situation because there are more parameters of the liquid absorber which are analyzed. Therefore, formula (4) has been modified by the following:

$$r = k_0 \times (\frac{4}{3})^{(\sum_{i=1}^{n} k_i f_i) + b} \quad (5)$$

where $f_i$ is the impact factor of the absorber. $n$ is the number of the impact factor and $k_i$ is the influence coefficient of each impact factor. b is the constant variable and is related to the shape of the liquid absorber.

For example, for the truncated cone-shaped absorber, there are three impact factors: the bottom diameter, the height and the top diameter. Therefore, $r$ can be described as:

$$r_1 = k_0 \times \left(\frac{4}{3}\right)^{(k_{11} \times d_1 + k_{12} \times d_2 + k_{13} \times h_{11}) + b_1} \quad (6)$$

where $d_1$, $d_2$ and $h_{11}$ are the top diameter, the bottom diameter and the height of the truncated cone-shaped absorber, respectively. Here, $k_0 = 0.005$, $k_{11} = 199.1$, $k_{12} = 164.8$, $k_{13} = 310.1$, $b_1 = 9.58$. Similarly, for the prismatic-shaped absorber, $r$ can be described as:

$$r_2 = k_0 \times \left(\frac{4}{3}\right)^{(k_{21} \times W_1 + k_{22} \times W_2 + k_{23} \times h_{21}) + b_2} \quad (7)$$

where $W_1$, $W_2$ and $h_{21}$ is the top width, the bottom width and the height of the prismatic-shaped absorber, respectively. Here, $k_{21} = 265.4$, $k_{22} = 186.1$, $k_{23} = 311.3$, $b_2 = 9.81$. To verify the correction of the equation (4), The first resonant frequency of the truncated absorber with different top diameter $d_1$ and the first resonant frequency of the prismatic absorber with different top width $W_1$ have been shown in Figs. 4(a) and 4(b), respectively.

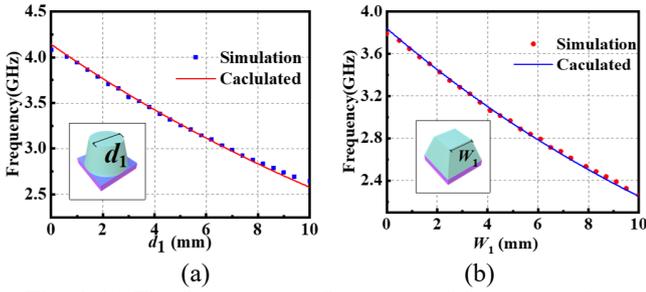

**Fig. 4.** (a) The first resonant frequency of the truncated cone-shaped absorber with different top diameters $d_1$. (b) The first resonant frequency of the prismatic-shaped absorber with different top widths $W_1$

From equation (2), we can find that $f$ becomes smaller when $r$ becomes bigger. When keeping the bottom diameter and the height of the truncated cone-shaped absorber unchanged, it can be known from equation (5) that with the increase of the top diameter, $r$ will become larger, resulting in a decrease in $f$. This is consistent with the conclusion put forward. From equations (6) and (7), we can find that $k_{21} > k_{11}$, $k_{22} > k_{12}$, $k_{23} > k_{13}$ and $b_1 \approx b_2$. Therefore, when the two shaped absorbers both have a similar size, $r_2 > r_1$. In other words, the first frequency of the prismatic-shaped absorber is lower than that of the truncated cone-shaped absorber. Last, the prismatic absorber has been chosen to realize good absorption property at low frequencies.

The unit cell of the absorber has been shown in Fig. 5(a). It is composed of three layers: the 3D container, the 45% ethanol layer and the metal. A prismatic 3D container has been designed to enclose the liquid. To facilitate the circulation of liquid between adjacent unit cells, a square liquid substrate that is used for slitting has been added to the liquid absorber. To achieve good impedance matching in a wide band, 45% ethanol is chosen to replace water. The reason has been introduced in [20]. The optimized parameters are shown as follows: $w_3 = 5$ mm, $w_4 = 8$ mm, $w_5 = 3$ mm, $w_6 = 7$ mm, $h_2 = 10$ mm, $h_3 = h_5 = 5$ mm, $h_4 = 9$ mm.

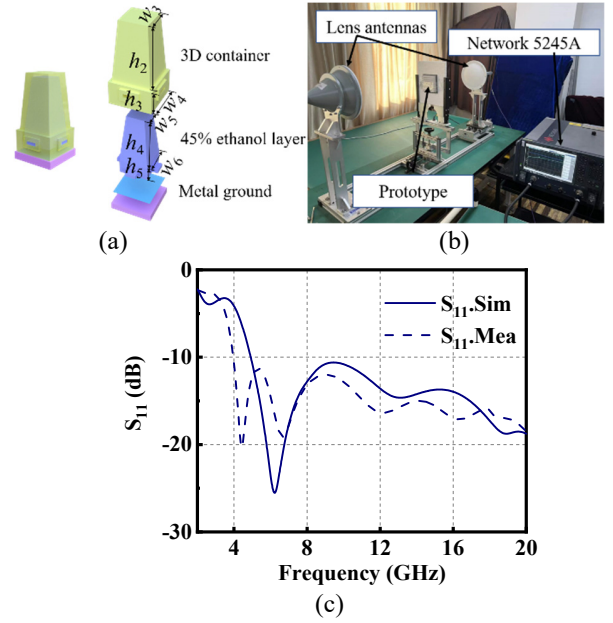

**Fig. 5.** (a) The unit cell of the liquid absorber. (b) Measurement setup of the $S_{11}$ of the absorber. (c) Simulated and measured $S_{11}$ of the liquid absorber.

The $S_{11}$ of the absorber has been simulated and measured, as shown in Fig. 5(c). It is noted that lens antennas and a 5245A vector network analyzer are chosen to measure the $S_{11}$ of the liquid absorber due to size constraints of the liquid absorber, as shown in Fig. 5(b). The absorption band of the absorber ranges from 4 GHz to 20 GHz and the absorption bandwidth is over 133%. The difference between measurement and experiment is affected by the following factors: 1) it is difficult to ensure that the container is completely filled with ethanol; 2) a small amount of misalignment of the 3D container and ethanol concentration is unavoidable in such a setup.

## III. GEOMETRY AND PROPERTIES OF THE INTEGRATED STRUCTURE

### A. Structure of the Versatile Integrated FP Antenna

The structure of the FP antenna is shown in Fig. 6. It is composed of the 3D container which is used to enclose the ethanol. The PRS is composed of patches on the top layer of the substrate and the square loop on the bottom. A microstrip patch antenna serves as the excitation source. The PRS and the metal ground of the antenna form an FP resonant cavity. It is noted that the height of the FP resonant cavity $h$ with container is lower than that without container because of the existence of the 3D container.

*B. Properties of the Versatile Integrated FP Antenna.*

To verify the properties of the FP antenna, the antenna has been fabricated and measured. The container structure which is used to enclose liquid is fabricated by 3D printed technology. The PRS and the microstrip patch antenna are fabricated by the PCB process. Figs. 7(a) and 7(b) show photographs of the fabricated FP antenna and PRS.

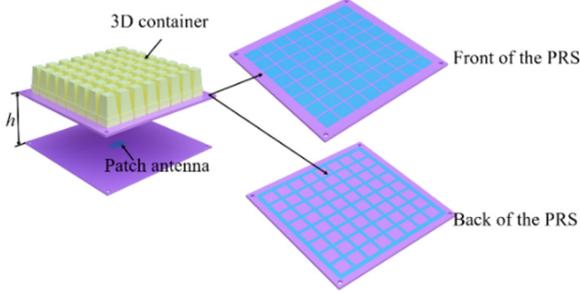

**Fig. 6.** 3D sketch of the FP antenna.

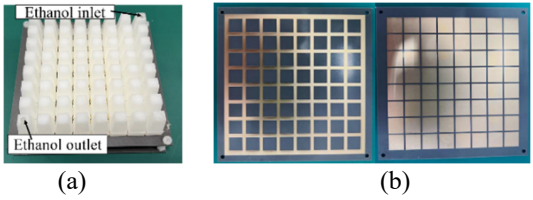

**Fig. 7.** Photographs of the fabricated (a) FP antenna and (b) PRS.

Simulated and measured $S_{11}$ of the FP antenna is shown in Fig. 8(a). The operating frequency of the antenna is about 10 GHz. The operating bandwidth is about 100 MHz. The gain of the antenna at 10 GHz is 19.7 dB, as shown in Fig. 8(b).

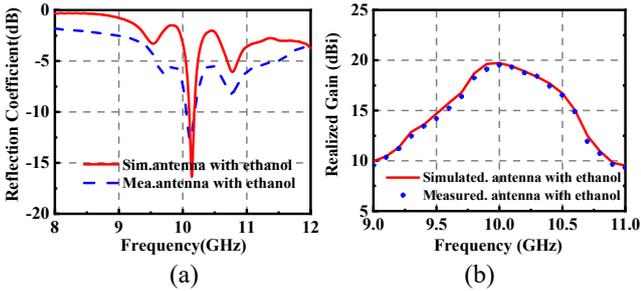

**Fig. 8.** Simulated and measured (a) $S_{11}$ and (b) realized gain of the FP antenna.

The radiation pattern of the FP antenna has been measured in a microwave anechoic chamber, as shown in Fig. 9. The E-plane and H-plane radiation patterns of the antenna have been shown in Figs. 10(a) and 10(b).

The monostatic RCS of the structure is shown in Fig. 11(a). It can be found that the band of RCS reduction is consistent with the absorption band of the absorber. When the frequency is lower than 4 GHz, the RCS of the two states of the structure keeps almost unchanged. When the frequency is higher than 4 GHz which is in the absorption band of the absorber, the RCS reduction is obvious. The RCS reduction of the structure with or without ethanol is shown in Fig. 11(b). The average RCS reduction is 28.4 dB. The RCS of the structure isn't measured because of the size constraints.

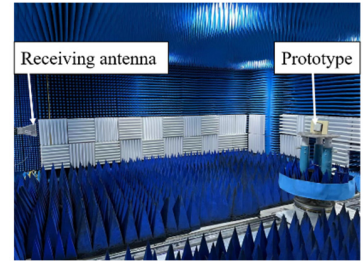

**Fig. 9.** Measurement setup of the radiation performance

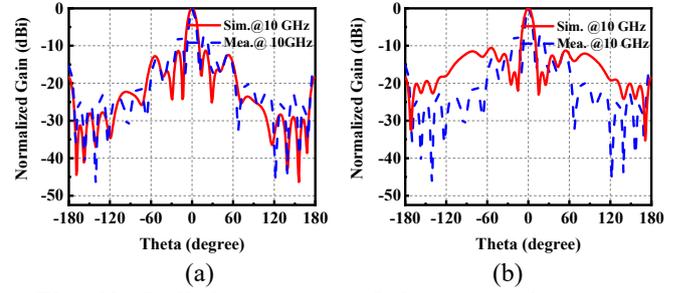

**Fig. 10.** Radiation patterns of the proposed antenna at 10GHz. (a) E-plane (b) H-plane

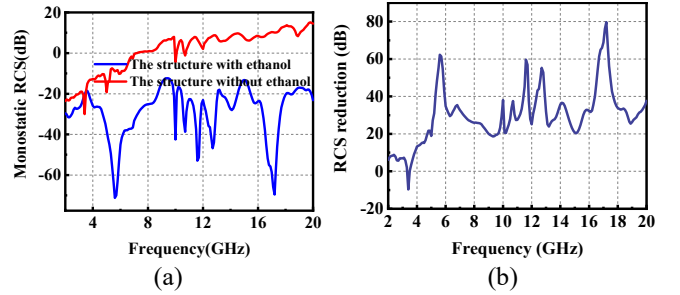

**Fig. 11.** (a) Monostatic RCS of the structure. (b) RCS reduction of the structure.

## V. CONCLUSION

A functionally reconfigurable integrated structure of FP antenna and wideband liquid absorber has been proposed in this paper. When the ethanol is injected, the structure acts like an absorber. The absorption band ranges from 4 GHz to 20 GHz and the absorption bandwidth is over 133%. When the ethanol is extracted, the structure acts like an FP antenna. The antenna operates at 10 GHz and the gain of the antenna is 19.7 dBi. The structure possesses reconfigurable RCS by injecting or extracting and this can adapt to the complex electromagnetic environment. The RCS reduction of the structure ranges from 4 GHz to 20 GHz and the average RCS reduction is 28.4 dB.